\documentclass[twocolumn]{aastex61}
\usepackage{graphicx}
\usepackage{multirow}
\usepackage{placeins}
\usepackage{color}
\usepackage{verbatim}
\usepackage{footnote}
\usepackage{soul}
\usepackage{amssymb}
\usepackage{amsmath}
\usepackage{hyperref}

\received{March 24, 2018}
\revised{May 9, 2018}
\accepted{May 11, 2018}
\submitjournal{PASP}

\begin{document}
\title{Estimating Sky Level}

\author[0000-0001-5706-0686]{Inchan Ji}
\author[0000-0001-5898-6499]{Imran Hasan}
\author[0000-0002-5091-0470]{Samuel J. Schmidt}
\author[0000-0002-9242-8797]{J. Anthony Tyson}
\affiliation{Department of Physics, University of California, One Shields Ave., Davis, CA 95616, USA} 
\email{tyson@physics.ucdavis.edu}

\begin{abstract}
We develop an improved sky background estimator which employs optimal filters for both spatial and pixel
intensity distributions. It incorporates growth of masks around detected objects and a statistical estimate of the flux
from undetected faint galaxies in the remaining sky pixels. We test this algorithm for underlying sky estimation
and compare its performance with commonly used sky estimation codes on realistic simulations which include
detected galaxies, faint undetected galaxies, and sky noise. We then test galaxy surface brightness recovery using
GALFIT 3, a galaxy surface brightness profile fitting optimizer, yielding fits to S\'{e}rsic profiles. This enables robust
sky background estimates accurate at the 4 parts-per-million level. This background sky estimator is more accurate
and is less affected by surface brightness profiles of galaxies and the local image environment compared with other
methods.
\end{abstract}


\section{Introduction}
\label{intro}
Detection and surface photometry of faint objects rely heavily on accurately 
estimating the underlying sky background flux. From scattered light originating 
from astronomical objects such as the Sun, the Moon, the Milky way, stars, and 
galaxies, to light pollution from the ground, there are many sources that 
contribute to the night sky surface brightness \citep{Roach:1973aa}. Therefore, 
all ground-based telescopes encounter the challenge of estimating and 
subtracting the night sky surface brightness ($\mu_{\rm sky} 
\simeq 21~{\rm mag~arcsec^{-2}}$ at typical ``dark'' locations) in order to 
access the far smaller flux levels characteristic of faint galaxy halos. Not 
only ground-based telescopes, but also space-based telescopes must tackle the 
issue of sky subtraction.  The sky surface brightness measured by the $Hubble~ 
Space~Telescope~(HST)$ is 1-2 mag arcsec$^{-2}$ fainter than the sky surface 
brightness measured by ground-based telescopes \citep{Trujillo:2016aa}, but 
contaminating flux sources remain: above the atmosphere zodiacal light, airglow 
from the Solar wind, and excitation of residual propellant gas from spacecraft 
all contribute to the sky background. 

The proper sky level can be different for detection of objects than it is for 
the optimal measurement of photometry. This is due to the fact that faint, 
unresolved and undetected objects underlie the object for which photometry is 
desired. This is true for stars as well as galaxies. Unbiased sky estimation has 
been attempted widely in the literature: prominent examples include 
FOCAS \citep{Tyson:1979aa}, DAOPHOT \citep{Stetson:1987aa}, 
SExtractor \citep{Bertin:1996aa}, SDSS Photo \citep{Lupton:2002aa}, 
GALFIT \citep{Peng:2010aa}, PyMorph 
\citep{Vikram:2010aa,Bernardi:2017aa,Fischer:2017aa}, 
and the LSST Data Management Stack 
\citep[LSST Stack hereafter,][]{Bosch:2018aa}, \citet{Huang:2018aa}, and \citet{Jenness:2015aa}. 
The problem of using biased sky background around detected objects is typically 
encountered on scales that are large compared with the point spread function (PSF), 
where the pixel counts from the object become indistinguishable from the sky 
pixel counts. Traditionally, detection of low surface brightness galaxies has relied on background 
sky estimation precision of one part in 10,000. Extreme dwarf galaxies in the 
Local Group have mean surface brightnesses as faint as $\sim32$ mag 
arcsec$^{-2}$ \citep[e.g.,][]{McConnachie:2009aa,Homma:2016aa}. 
Thus, accurate surface brightness measurements would require a sky unbiased 
at a level of $\sim34$ mag arcsec$^{-2}$, or about 6 parts-per-million (ppm) of 
the typical $R$-band sky level. 

Current and upcoming surveys such as the Dark Energy Survey 
\citep[DES;][]{Flaugher:2005aa}, the Large Synoptic Survey Telescope 
\citep[LSST;][]{LSSTSciBook:2009}, and the Hyper Suprime-Cam (HSC) survey 
\citep{Aihara:2018aa} are likely to reveal new aspects of galaxies 
as low surface brightness (LSB) objects \citep{Ivezic:2008aa,Robertson:2017aa}.
The discovery space is large: LSB features can exist on scales of arcseconds to 
many arcminutes, spanning the majority of faint galaxies at high redshift to 
more nearby LSB galaxies. Tidal tails have already been detected at surface 
brightness levels of $\sim30$ mag arcsec$^{-2}$ 
\citep[e.g.,][]{Martinez-Delgado:2010aa,van-Dokkum:2014aa} 
and surely exist at lower levels. A relatively unexplored area is the 
ultra-low surface brightness morphology and tails over a wide range of angular 
scales at levels of 31-32 mag~arcsec$^{-2}$. Discoveries are likely at even 
fainter levels of surface brightness still, which may become accessible in upcoming 
deep field observations such as the LSST Deep Drilling Fields (hereafter LSST 
DDFs), which are expected to achieve a coadded 5$\sigma$ depth of 
$\sim$29 magnitude in the $r$-band filter \citep{LSSTSciBook:2009}. 

Proper sky background estimation is an important tool for studying galaxy 
formation and evolution, and even more important when estimating galaxy types 
based on surface brightness profiles. However, it has been challenging to 
calculate the correct value of sky background. Many automated photometry 
programs estimate a biased sky background \citep{Bosch:2018aa,Huang:2018aa,Jenness:2015aa}. 
Previous techniques typically mask detected 
objects and use the remaining pixel values to estimate the background level; 
however, sky estimates are generally biased high because pixels in the outskirts 
of detected galaxies survive the masking processes, and undetected low 
signal-to-noise sources contaminate the background. Accurate sky estimation is 
thus a prerequisite for photometric studies of faint objects (e.g., LSB 
galaxies or low-level features around galaxies).

As mentioned above, at these low levels of surface brightness a sufficiently 
accurate model of camera scattered light must be used for each exposure. 
Indeed on a wide range of angular scales the sky surface brightness will be 
dominated by scattered light from bright stars.  Such modeling is beyond the 
scope of this paper, instead we focus on the challenge of sky bias introduced by 
the detected object's faint outer halo and by the high density of undetected 
galaxies. Thus, suppressing sky bias from these two known effects is a 
necessary but not sufficient condition for ultra-low surface brightness 
photometry \citep{LSSTSciBook:2009}.

Sky background estimation directly impacts astronomical object detection. 
Detection of both stars and galaxies requires accurate characterization of the 
sky background. As an example, photometry of faint stars whose surface 
brightnesses are very near that of the sky is described in 
\citet{Stetson:1987aa}. Typically, detection algorithms begin by marking a 
collection of CCD pixels as belonging to an astrophysical source if they are 
above some threshold, usually after convolution with a spatial filter optimized 
for some angular scale. Calculating the flux due to this source (and crucially, 
this source alone) requires that we quantify the flux those CCD pixels would 
have in the absence of contaminants, e.g. the sky background. In virtually all 
sky surveys the flux from unresolved, undetected faint galaxies form a component 
of this background sky.  However, their number and luminosity distribution is 
known statistically from existing deep surveys. Compilations of deep imaging 
data provide the number of galaxies as a function of magnitude up to $m
\simeq 30$ for various astronomical filters. For example, 
\citet{Metcalfe:2001aa} showed that the galaxy count slope in 
$R$-band is $d({\rm log}N)/dm_{\rm R}$ $\sim$ 0.37 for 20 
$\lesssim m_{\rm R}\lesssim$ 26 and becomes shallower for 26 
$\lesssim m_{\rm R}\lesssim$ 30. This complete galaxy number count was achieved  
by compiling a number of observations from both ground-based and space 
telescopes. Data obtained by a single survey rarely satisfy both a large field 
of view and a very deep image depth. In deep imaging covering a sufficiently 
large area (to avoid sample variance) one can statistically expect the same 
galaxy number counts from any observation at that wavelength. We may thus adopt 
the well measured mean number of faint undetected galaxies which are responsible 
for biased sky estimates if the imaging is sufficiently deep. 

The idea of correct sky estimation over all angular scales is actually an 
ill-posed problem. The proper background sky for barely resolved galaxies at 
high redshift is, in principle, quite different from the correct sky level for 
large angular scale LSB features. Indeed, the flux from barely resolved 
galaxies sits on top of the fainter, larger, angular scale flux associated 
with arcminute scale LSB extragalactic features, which in turn sits on top of 
the starlight reflected by Galactic cirrus, the zodiacal light, the night sky 
surface brightness caused by atmosphere emission, and scattered light from 
bright objects in the camera and the atmosphere. Thus there could be a separate 
sky estimate appropriate for each of the different morphological classes of LSB 
objects. To make the problem tractable a multi-component sky model must be 
built. 

The sky model, in principle, can be built using knowledge of the camera and 
telescope system, locations of bright objects, observational data, and 
statistical summaries of faint galaxy counts from ultra deep images like the 
HST. The first step is detecting all objects above a position-variable local 
sky estimate and masking them. The remaining pixels still contain flux from both 
undetected galaxies and the faint outer isophotes of the masked detected 
galaxies which, if left uncorrected,  gives an over-estimate of the sky level around 
compact objects.  Because of this, fitting the remaining ``sky" pixels with a 
Gaussian profile, as if it were pure Poisson noise, is incorrect; the 
distribution of remaining pixels would follow a Gaussian if the pixels contain 
only the true sky. However, the real distribution has a tail of positive pixels 
due to the two contributors mentioned above. While 3$\sigma$ clip and/or 
one-sided Gaussian fitting improves the estimate, these approaches are arbitrary 
and lead to a small positive sky background bias \citep{Robertson:2017aa}. 

Using the known statistical faint galaxy counts beyond the detection limit 
together with growing masks around detected objects by a defined amount scaled 
by total flux help significantly in making these corrections. Indeed, both 
biases must be removed if the sky is to be  correctly estimated at the 
sub-percent level. This entire process is recursive on every angular scale where 
there are important sky components. In this paper we focus on the more tractable 
task of estimating the sky level in the generic case of the extragalactic sky 
superposed on a slowly varying foreground, thus focusing on the $\sim$few 
arcsecond scales associated with typical faint galaxies. To explore the effect 
of modified masking and accounting for undetected galaxies in a controlled way, 
we develop a set of simulated images with known inputs and properties.

We begin by describing our image simulations in Section \ref{GALSIM}. 
In Section \ref{Object Detection} we outline the methodology for creating detection 
and measurement catalogs with SExtractor, where we detail the software 
specific settings used in this analysis. Section \ref{skyestimators} continues with 
a discussion of two widely used sky estimators and the techniques they employ. 
In Sections \ref{newestimationtechnique} and \ref{stack+inchan}, we present our 
new sky estimation technique which deals with biases current sky estimators 
suffer. In Section \ref{sersicest} we examine the importance of accurate sky 
background estimation in the fitting of galaxy surface brightness profiles. We 
close this work in Section \ref{discussion} with a discussion of the difficulties 
of correctly estimating sky backgrounds, the effectiveness of our algorithm to 
overcome them, and prospects for future directions.

\begin{deluxetable*}{cccc}[ht]
\tablecaption{Parameters used in GALSIM simulations}
\tablehead{
\colhead{Model} & \colhead{Parameter}& \colhead{Simulation} & \colhead{Range}
}
\startdata
\multirow{4}{*}{\shortstack[c]{PSF (Moffat)}} 
&FWHM [arcsec] &\multirow{4}{*}{All}& 0.8 $\le \alpha \le$ 1\\
&slope, $\beta$ && 2.9 $\le \beta \le$ 3.2\\
&ellipticity, $e$ && 0 $\le e \le$ 0.15 \\
&position angle, $\theta$ [degree] 
&& $0^\circ\le \theta \le$ 180$^\circ$\\
\hline
\multirow{2}{*}{{Sky (DLS Depth)}} 
& sky level, $\mu_{\rm sky}$ [ADU~pixel$^{-1}$] 
& All &$\mu_{\rm sky}=$ 3240\\
& sky noise, $\sigma_{\rm sky}$ [ADU~pixel$^{-1}$] 
& All &$\sigma_{\rm sky}=$ 12.73\\
\hline
\multirow{2}{*}{{Sky (LSST DDF Depth)}} 
& sky level, $\mu_{\rm sky}$ [ADU~pixel$^{-1}$] 
&  \multirow{2}{*}{Uniform dist.}$^{\rm a}$ 
&$\mu_{\rm sky}=$ 1000\\
& sky noise, $\sigma_{\rm sky}$ [ADU~pixel$^{-1}$] 
& &$\sigma_{\rm sky}=$ 0.1\\
\hline
\multirow{8}{*}{Galaxy} 
&\multirow{2}{*}{magnitude in $R$ band, $m_{\rm R}$} & Uniform dist. 
& 19 $\le  {m_{\rm R}}\le$ 25\\
&&Random dist.& 19 $\le  {m_{\rm R}}\le$ 29 \\ 
&\multirow{2}{*}{half-light radius, $R_{e}$ [arcsec]} 
& Uniform dist.& 0.3 $\le R_{e} \le$ 2.5\\
&&Random dist.& 0.1 $\le R_{e} \le$ 2.5\\
&{S\'{e}rsic index, $n$} & All & 0.5 $\le n \le$ 5\\
&\multirow{2}{*}{ellipticity, $e$ } & \multirow{2}{*}{All} 
& 0 $\le e \le$ 0.6 (for $n$ $\le$ 2.5)\\
&&  & 0 $\le e \le$ 0.3 (for $n$ $\ge$ 2.5)\\
&position angle, $\theta$ [degree] &All 
& $0^\circ\le \theta \le$ 180$^\circ$\\
\enddata
\tablenotetext{}{\textbf{Note.} We assume that 
the point-spread-function follows a Moffat profile and a sky surface brightness of 
$\mu_{\rm sky} =$ 21 mag~arcsec$^{-2}$. We generated and co-added 20 images to 
increase the signal-to-noise, as in the DLS. The half-light radius of each galaxy 
scales with its magnitude, as in the observed data.}
\tablenotetext{}{$^{\rm a}$ Only the uniform distribution is simulated to LSST DDF depth. This is done 
to investigate the effect of lower sky noise on the galaxy surface brightness fits.}
\label{tbl:galsim} 
\end{deluxetable*}

\section{GALSIM: Galaxy Image Simulator}
\label{GALSIM}
We use GALSIM \citep{Rowe:2015aa} to generate galaxy images and sky background. 
Galaxy and PSF parameters are chosen to be similar to 
those observed in the $R$-band imaging data of the Deep Lens Survey 
\citep[DLS;][]{Wittman:2002aa}. In all, three images are simulated in which we vary 
the sky level, sky noise, galaxy placement, and magnitude distribution. Each 
simulation is 8000 $\times$ 8000 pixels in area. In Table \ref{tbl:galsim} we 
list the parameters used in our simulations.

\begin{figure*}[ht]
\centering
\includegraphics[width=0.9\textwidth]{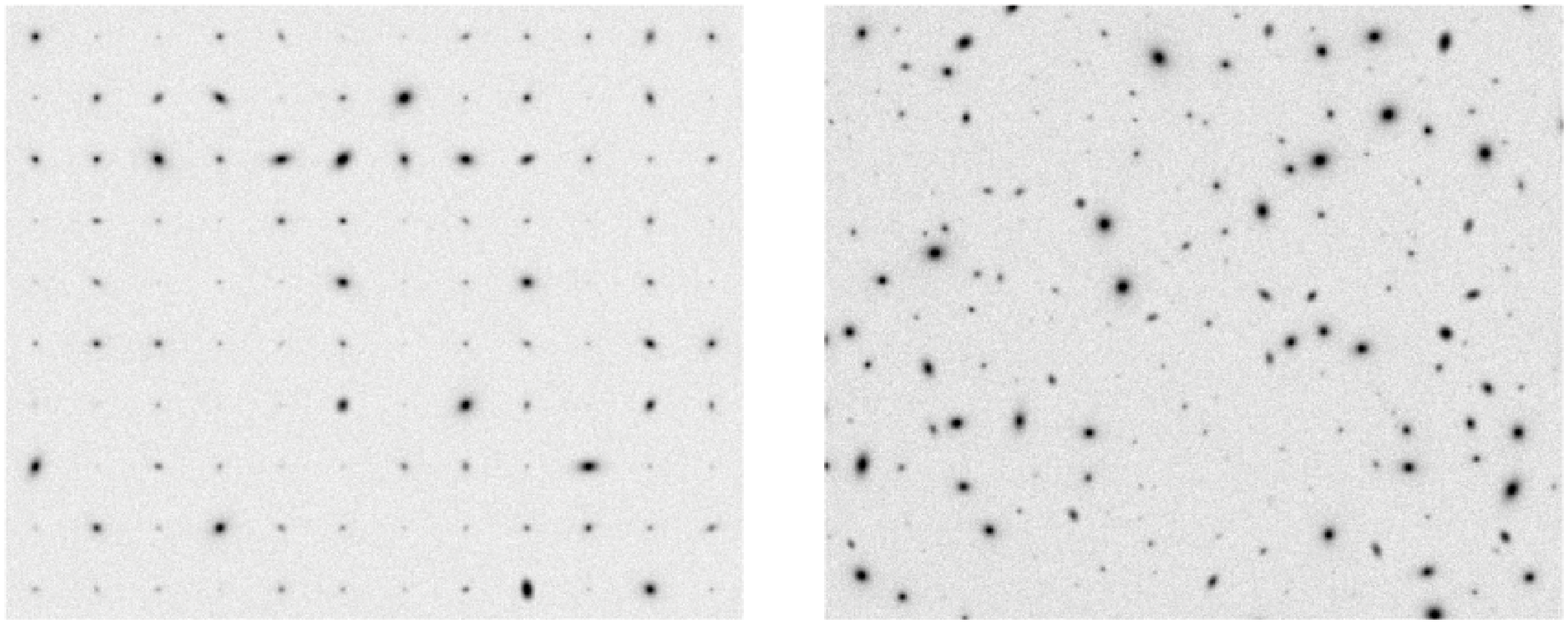}
\caption{
Representative portions of our simulated galaxy images generated by GALSIM 
to DLS depth.  
Left: the uniform distribution simulation. Galaxies are evenly spaced, and 
all galaxies are above the detection limit. This allows us to examine the effect 
of growing detection masks on background estimation, independent of 
contributions from undetected background sources. 
Right: the random distribution simulation. 
Galaxies are placed at random on the image. A population of magnitude in R 
band, $m_{\rm R}~=~$25.5, and fainter are below the detection threshold. 
This combination of simulations enables us to test the effects of growing 
detection masks around galaxies both on isolation and in combination with the 
effects arising from the inclusion of undetected galaxies.}
\label{GALSIM:IMG}
\end{figure*}

In order to isolate the effect of detection 
masks from that of undetected faint galaxies, in our first simulation 
(hereafter referred to as the ``uniform distribution simulation'') 
10,000 galaxies are evenly spaced in the image in a grid pattern, and are not 
surrounded by {\it any} other galaxies inside the sky analysis areas. This ensures the 
galaxies are well separated and do not contaminate their neighbors with stray 
flux. As discussed earlier, extended features in detected galaxies often 
`bleed' beyond the detection mask, contaminating the sky background estimate. 
As we will show in further detail in Section \ref{SExSky}, galaxies in the uniform 
distribution simulation are all detected, meaning there will be no sky 
background contribution from unresolved or undetected sources. Having completely 
detected all simulated galaxies, well localized sources will enable us to directly test 
the impact of growing detection masks, independent of any effects from undetected 
sources in more realistic images. 
The number count as a function of magnitude in R band follows a power law with 
a shallow slope of $d({\rm log}N)/m_{\rm R}$ = 0.1 over the magnitude 
range of 19 $<m_{\rm R}<$ 25 for galaxies in this ``uniform distribution" 
simulation. 
We choose a shallower slope for the number counts to ensure that our simulated 
population includes a balanced mix of bright and faint galaxies in our sample of 
10,000.  A slope of $d({\rm log}N)/dm_{\rm R} =$ 0.1  yields a sample that 
contains adequate bright galaxies to test our algorithms while minimizing problems 
due to flux overlaps that may occur from two neighboring bright galaxies on our 
simulation grid.

The second simulation (hereafter referred to as the ``random distribution 
simulation") contains 793,116 galaxies which are randomly placed in position 
over the image. As a consequence of their more realistic placement, galaxies 
may contribute flux to neighboring profiles. Importantly, the random 
distribution simulation includes a much larger number of  galaxies, with an $n(m)$ 
distribution that extends to a much fainter magnitude limit of 
$m_{\rm R}$ = 29. Magnitudes of galaxies in the random distribution 
simulation follow a power law with a slope of 
$d({\rm log}N)/dm_{\rm R} =$ 0.4. The fainter galaxies with 
$m_{\rm R}>$ 25.5 are mostly not detected by the detection algorithms tested 
in this paper. 
We call the flux from these undetected galaxies the extragalactic background 
(EBL). The EBL will contaminate both the estimate of the sky background, and 
the flux of nearby detected galaxies.  These undetected galaxies will be 
important in Section \ref{newestimationtechnique}. Taken together, the two 
simulations enable us to test the effect of mask growth in isolation in the 
uniform distribution simulation, and the joint effects of mask growth and 
unresolved galaxies in the random distribution simulation. 

In Figure \ref{GALSIM:IMG} we show a small portion of each of the first 
two simulations generated by GALSIM. The simulated images are intended to 
reflect the observing 
conditions in the DLS, with 20 co-added 900 s R band exposures on a 
4m telescope. We constructed 20 simulated images with different random seeds and 
co-added the images for each simulation. In doing so, we increase the 
signal-to-noise ratio and limiting magnitude.  The simulated sky in both the uniform and random distribution simulations properly 
emulates conditions of the DLS. The sky surface brightness is 
$\mu_{\rm sky}~=~21~{\rm mag~arcsec^{-2}}$ (3240 ADU~pixel$^{-1}$), which 
corresponds to a single exposure time of 900 s in DLS R band with 
Poisson noise. We also assume that the sky surface brightness is spatially 
flat in the uniform and random distribution simulations. The 
root-mean-square value of sky background in the co-added image is reduced by a 
factor of the square root of the number of co-added images and becomes 
$\sigma_{\rm sky}$ = 12.73 ADU~pixel$^{-1}$.  

The range of signal-to-noise ratio is 5 $\lesssim$ S/N $\lesssim$ 300 for 
19 $\leq m_{\rm R} \leq$ 26 at DLS depth. 
Each model galaxy follows a single S\'ersic profile 
with index ranging from 0.5 to 5. The PSF profiles in the DLS are 
broader than a Gaussian, and are well described by a Moffat profile 
\citep{Moffat:1969aa} which is given by:
\begin{equation}
{\rm PSF}(R) = \frac{\beta-1}{\pi \alpha^2}
\bigg[ 
1 + \bigg(\frac{R}{\alpha} \bigg)^2
\bigg]^{-\beta}, 
\end{equation}
where $\alpha$ is the scale length, and $\beta$ is the slope of the profile. 
To match the DLS observations, we use Moffat profiles of 
$0.8 <\alpha [{\rm arcsec}]< 1$ and $2.9 < \beta < 3.2$. Under these parameter 
ranges, Moffat PSFs are randomly distributed in the entire image for both 
simulations.

To investigate effects of noise, it is informative to simulate deeper data 
with higher signal to noise for the target galaxies. To this end, for our third image simulation, we regenerate 
the uniform distribution simulation, this time to LSST DDF depth, in addition to the simulations 
to DLS depth. 
Anticipated typical observing conditions for the LSST DDF are as follows: 10,000 co-added 
15 s R band exposures on a 6.7 m (effective aperture) telescope. 
The sky surface brightness of $\mu_{\rm sky}~=~21~{\rm mag~arcsec^{-2}}$ is 
estimated as 1000 ADU pixel$^{-1}$. 
The root-mean-square value of sky background in the co-added image is  
$\sigma_{\rm sky}$ = 0.1 ADU~pixel$^{-1}$.  
Simulating 10,000 simulations with varying random seeds requires a substantial 
amount of computing time. To impart realistic sky noise fluctuations in our LSST 
DDF depth image, we take the following steps instead: we re-normalize pixel values 
of the uniform distribution simulation without sky noise in DLS depth 
to meet the observing condition for the LSST DDF. We subsequently add the 
Poisson noise of $\sigma_{\rm sky}$ = 0.1 ADU~pixel$^{-1}$ to each pixel.

\section{Object Detection}
\label{Object Detection}
The correct sky level to be used in detection and photometry can differ. 
The proper sky level for object detection is the sky underlying all objects, 
bright and faint. This is true even though the faintest objects are generally 
not detected and form an unresolved extragalactic background. 
Any detection algorithm should use the true underlying sky after EBL subtraction. 
However, current algorithms are not sensitive at the levels discussed above. 
Developing a new detection algorithm which takes full advantage of the high precision 
sky estimates is beyond the scope of this paper. For the purposes 
of the inter-comparisons in this work we use SExtractor. SExtractor 
\citep{Bertin:1996aa} is an automated catalog builder used to identify and 
measure various properties of astronomical objects on a CCD image. We run 
SExtractor using a detection threshold of 
\verb+DETECT_THRESH+ = 0.5$\sigma$ where $\sigma$ is the root-mean-square 
sky noise in the entire image, a minimum detection area of 6 pixels, and the number 
of deblending sub-thresholds of \verb+DEBLEND_NTHRESH+ = 10 with a deblending 
contrast of \verb+DEBLEND_MINCONT+ = 0.0001. The images are filtered through 
a 5 pixel $\times$ 5 pixel Gaussian convolution kernel with FWHM = 3 pixels.
A mesh of 80 pixels $\times$ 80 pixels is used to estimate sky background 
for all identified galaxies by SExtractor. A \verb+PHOT_FLUXFRAC+ = 0.5 is
used to estimate the half-light radius for each galaxy. The numbers of 
cataloged galaxies are 10,000 and 111,409 for uniform and random 
distributions, respectively. As in all current object detection algorithms, 
SExtractor fails to detect faint galaxies of low signal-to-noise.

\clearpage
\section{Background Estimation with Three Sky Estimators}
\label{skyestimators}
In this section, we compare sky background 
values estimated by various methods on our simulated images. Because we 
have \textit{a priori} knowledge of the true underlying sky brightness that was input 
into our simulations, we can directly assess the accuracy of these estimators by 
comparing their results with the truth. To estimate sky background around 
each of our galaxies, we first run SExtractor to construct a detection catalog of 
galaxies. By running SExtractor, we obtain SExtractor's sky background estimate 
at the position of each galaxy. The SExtractor catalog is used as input to GALFIT, 
which provides a second catalog of background estimations. An overview of the 
background estimation procedures and our 
parameters used in SExtractor and GALFIT follows in the next two subsections.

Motivated by the strengths and weaknesses observed in these methods, 
we develop a new scheme for estimating the local sky background around 
galaxies. Finally, we couple our local sky estimation technique with a global polynomial 
background model in each of the simulations to obtain accurate sky background 
estimates with high precision.

\subsection{SExtractor sky estimation}
\label{SExSky}

SExtractor provides a $local$ sky background estimate. In SExtractor this quantity is 
estimated by performing an iterative 3$\sigma$ clip of the pixel values within a 
user-specified mesh grid that covers the image. We use a mesh of 80 
pixels $\times$ 80 pixels to estimate sky background.
Sextractor considers the cell to be ``non-crowded" if $\sigma$ drops by less than 
0.2$\sigma$ per clipping iteration, and crowded otherwise.  
Based on these two cases the sky background is given by:
\begin{equation}
{\rm sky} = \begin{cases} 
{\rm Mean}  & \sigma_i - \sigma_f \le 0.2 \sigma_i \\
2.5 \times {\rm Median} - 1.5 \times {\rm Mean} &{\rm otherwise}
\end{cases}
\end{equation}
where $\sigma_i$ and $\sigma_f$ are the standard deviations of 
the pixel values in a mesh before and after the 3$\sigma$ clip, respectively.

\subsection{GALFIT sky estimation}
\label{GALFITSky}
GALFIT \citep[version 3;][]{Peng:2010aa} is a two-dimensional model fitter 
designed to model multiple categories of astronomical objects. In the course of 
measuring a model galaxy GALFIT estimates sky background, which is subtracted in order 
to find the best-fit parameters for a functional model on a CCD image. GALFIT 
estimates the sky background at the object's centroid as follows: 
\begin{equation}
\begin{aligned}
{\rm sky} (x_0,y_0) = &~{\rm sky} (x_c,y_c) +  (x_0-x_c)
\frac{d{\rm sky}}{dx}\\ 
&+  (y_0-y_c)\frac{d{\rm sky}}{dy},
\end{aligned}
\end{equation}
where $(x_0, y_0)$ is the centroid of the object in pixel coordinates, 
$(x_c, y_c)$ is the center of an image cutout and $d{\rm sky}/{dx}$ and 
$d{\rm sky}/{dy}$ are gradients of sky background in x and y directions, 
respectively.

Care must be taken in choosing the background estimation parameters for 
GALFIT, 
particularly when choosing an image cutout size. If the size is too small, the image 
cutout does not include enough sky pixels to make an accurate estimate. 
However, too large of an image cutout not only requires 
expensive computational resources, but also results in inaccurate estimation 
of the sky background due to the increasing number of 
undetected galaxies \citep{Barden:2012aa,Vikram:2010aa}. In our study, we adaptively 
choose the width $w$, and height $h$, of the image cutout centered on each galaxy's 
position. It is crucial that the cutout does not truncate the faint tails of 
galaxies. To ensure this is the case, we adopt a 
scheme to conservatively estimate the radius at which 
the galaxy profile reaches a surface brightness 
of $\mu$ = 30 mag arcsec$^{-2}$, $R_{\rm 30}$. We assume all galaxies are 
described by an $n$ = 4 S\'ersic profile and use the half-light radius as measured 
by SExtractor as the profile's half-light radius. After the mock profile is constructed, 
$R_{\rm 30}$ can be readily calculated for each galaxy. The width $w$, and height 
$h$, of the image cutout are then defined as:
\begin{eqnarray}
w &= f_{\rm img} R_{\rm 30} (|\cos\theta| + (1-e)|\sin\theta|),\\
h &=  f_{\rm img} R_{\rm 30} (|\sin\theta| + (1-e)|\cos\theta|),
\end{eqnarray}
where $\theta$ is \verb+THETA_IMAGE+, $e$ is \verb+ELLIPTICITY+, 
and $f_{\rm img}$ is a free parameter to set 
the optimal size of an image cutout. We empirically determine that a value of 
$f_{\rm img} = 2$ results in a sufficient number of background sky pixels to 
determine an accurate estimate of the sky background.
We use this same image cutout for galaxy surface brightness 
profile fitting in Section \ref{sersicest}.

\subsection{New Sky Estimation Technique}
\label{newestimationtechnique}
Sky estimation methods which use a sample of local image CCD pixels to 
estimate the background level at the position of a galaxy can suffer a high 
bias from two factors: flux from the outer tails of galaxy profiles which 
extend beyond their respective masks, and flux from undetected (and 
hence completely unmasked) faint galaxies that reside in the image pixels 
used to estimate the sky. To deal with these two effects, we develop a 
new sky estimation technique. This technique employs a two-filter 
estimator, one spatial and one statistical, to minimize the contribution 
from pixels coming from the unmasked outskirts of galaxy profiles and 
from unidentified objects. Our method consists of three high level steps 
and ultimately yields an estimation of the sky level at the positions of 
detected galaxies. A flow chart describing the overall
process of our sky estimation is shown in Figure \ref{Flowchart}.

In the spatial filtering step, we create an updated object mask for each 
cataloged source. These new masks more effectively exclude flux from the 
extended tails of galaxy profiles, which previously contaminated the pixels 
used to estimate the sky background. The procedure 
for creating new masks was calibrated on the uniform distribution 
simulation, where galaxies are laid down on a regular grid, with no 
neighboring galaxies inside the cutout area. The masks are generated 
by first creating a mock one-dimensional S\'ersic profile for each source. 
Pixels are then masked out if their positions satisfy:
\begin{equation}
\begin{aligned}
&{\rm C_{xx}}(x-x_c)^2 + {\rm C_{xy}}(x-x_c)(y-y_c) \\
&+ {\rm C_{yy}}(y-y_c)^2< (R_{30}/a_{\rm rms})^2  
\end{aligned}
\end{equation}
where $a_{\rm rms}$ is the 2nd moment along the semimajor axis 
(\verb+A_IMAGE+),  
(${\rm C_{xx}}$, ${\rm C_{xy}}$, ${\rm C_{yy}}$) is the object ellipse parameter
(\verb+CXX_IMAGE+, \verb+CXY_IMAGE+, and \verb+CYY_IMAGE+, 
respectively) measured by SExtractor, ($x_c$, $y_c$) is the centroid 
of a galaxy, and $R_{\rm 30}$ is the cutout radius (see Section \ref{GALFITSky}).
We find that 
using a fainter surface brightness for our masks does not significantly 
change the remaining pixel statistics. Additionally, the measured 
SExtractor ellipse parameters are used to assign orientation angles and 
ellipticities to the masks. We stress that these masks are not meant 
to perfectly model the profiles of galaxies, but rather effectively mask 
out their flux. Using an $n$ = 4 S\'ersic parameter is sufficient to 
mask out galaxies which are best described by 
$0.5\leq\,n\,\leq\,4$ S\'ersic profiles. 

\begin{figure}
\centering
\includegraphics[width=0.47\textwidth]{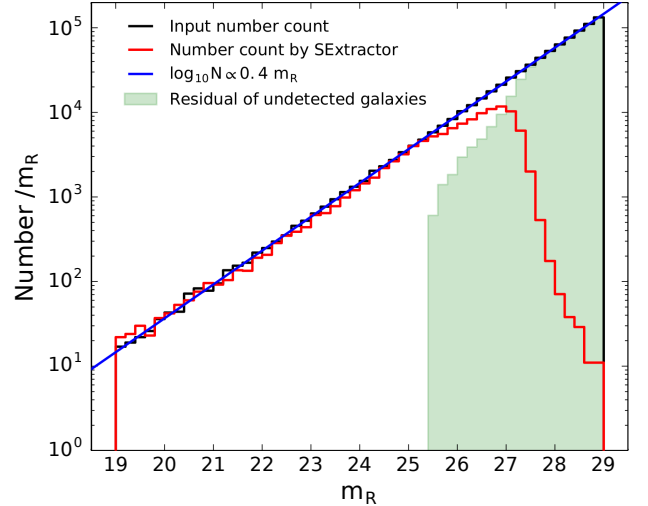}
\caption{
Galaxy number counts as a function of magnitude in the random 
distribution simulation: 
input number counts (black), power law with a slope 
of $d({\rm log}N)/dm_{\rm R}$ = 0.4 (blue), measurement by SExtractor 
(red), and residual counts (sea-green). The residual counts with 
$m_{\rm R}~>$ 25.5 
are used to model extragalactic background light of unmasked 
pixels.  We assume that the flux from these unresolved galaxies is 
uniformly distributed over the sky, which enables us to determine 
a background flux to subtract in our improved sky estimate.
}
\label{GalaxyCount}
\end{figure}

We then proceed to the statistical filtering step to estimate and subtract 
the flux contribution from undetected faint galaxies which make up the EBL. 
To model the EBL the number counts of undetected galaxies in the entire 
field and their flux must be considered. In the uniform distribution 
simulation, all galaxies are detected by SExtractor, so accounting for 
undetected EBL galaxies is unnecessary. In the random distribution 
simulation, however, SExtractor fails to detect and  catalog some galaxies 
at a true magnitude fainter than $m_{\rm R}\,\sim$  25.5, as the 
signal-to-noise for these galaxies approaches the user-set detection 
threshold. In Figure \ref{GalaxyCount} we show the galaxy number counts for the 
random distribution simulation. The blue curve shows the histogram of true 
magnitudes for the objects in the random distribution simulation, which 
is an excellent match to the the input number-counts slope that was 
used to generate the mock galaxies, shown in black. The red histogram 
shows the actual number of detected objects, and the 
filled sea-green histogram indicates the number of objects not detected 
by SExtractor, the difference of the red and blue histograms. 
We exclude a small number of galaxies with observed magnitude 
$m_{\rm R} <$ 25.5 when modeling the EBL; residuals in the observed 
number counts compared to the input number counts for $m_{\rm R} <$ 25.5 
galaxies are due to small measurement errors and the effects of blending. The total 
number of undetected galaxies with $m_{\rm R} >$ 25.5, which make up the EBL, is 
679,753. Since the number of EBL galaxies is large, for simplicity we assume that the 
galaxies are uniformly distributed in the field.

We use the residual in the observed number counts of galaxies as a function of 
magnitude compared to their expected value to calculate the total flux of all 
undetected galaxies. Because we assume EBL galaxies are uniformly distributed 
across the field, we also assume the total EBL flux is uniformly distributed as well. 
As a result, we obtain an estimate for the EBL flux per pixel for our simulated data, 
$\mu_{\rm EBL}$= 0.898 ADU~pixel$^{-1}$, by 
simply dividing the total EBL flux by the number of unmasked pixels in 
the image. 
This `pedestal' level of flux is then subtracted from each unmasked image pixel 
to mitigate the effects of the EBL in background estimation.
To examine whether the method is sensitive to the exact cutoff in the simulated faint 
galaxies, we test our method for EBL estimation on an additional image simulation 
where galaxies are generated up to $m_{\rm R}$ = 31 with a simple power-law of 
$d({\rm log}N)/dm_{\rm R} =$ 0.4. We find a background consistent with 
the added flux from the $29\,\leq\,m_{\rm R}\,\leq\,31$ galaxies: the EBL in this case 
is $\mu_{\rm EBL} = $ 1.670 ADU~pixel$^{-1}$ while the median pixel value is 
$\mu_{\rm median} = $ 1.593 ADU~pixel$^{-1}$. Thus, the method is not sensitive 
to the EBL faint end cutoff beyond 30 mag~arcsec$^{-2}$. 
Statistical galaxy counts as a function of magnitude are now 
complete to $m_{\rm R} \simeq$ 30 \citep{Metcalfe:2001aa}. 
The slope of the galaxy count at $m_{\rm R} \simeq 29$ becomes 
so shallow that the EBL from the galaxies with $m_{\rm R} > 29$ decreases rapidly 
\citep{Tyson:1995aa}. Because of this, there is little difference in the EBL 
estimates even though we simulate galaxies following the real galaxy counts. 
It is therefore safe to simulate galaxies with $m_{\rm R} < 29$.

Lastly, we measure local sky background estimates for each galaxy. 
We select an image cutout centered 
on the centroid of each target galaxy. The initial width and height of the 
images are 15 $R_{e} $, where $R_{e}$ is 
\verb+FLUX_RADIUS+ with \verb+PHOT_FLUXFRAC+ = 0.5 in our SExtractor 
catalog. If the number of unmasked pixels is less than 4000 or the width 
(or height) is less than 80 pixels, we iterate by increasing the width and 
height with an increment of 10 pixels. Once the number of unmasked 
pixels residing in the image cutout is greater than 4000, the mean of the unmasked 
pixels is calculated and used as an estimate of the local sky value for the center of 
the image cutout.

\begin{figure*}
\centering
\includegraphics[width=0.9\textwidth]{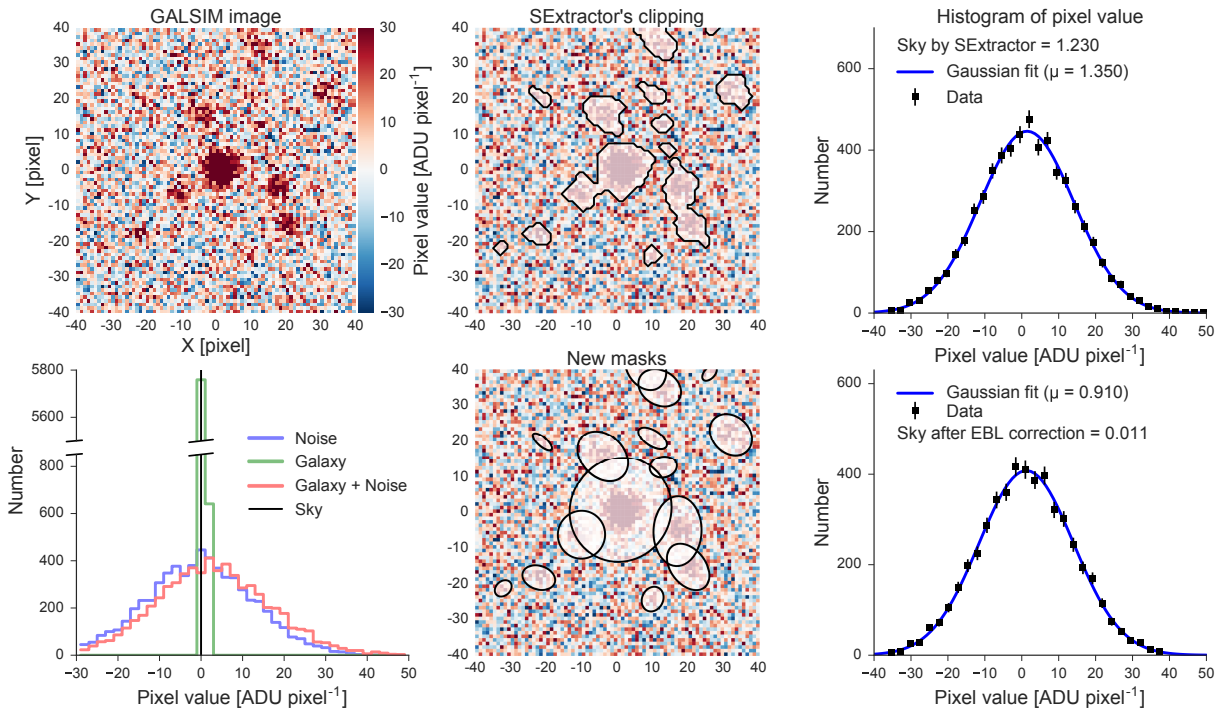}
\caption{
Comparison of sky background estimation: SExtractor vs. our new method. Width and 
height of image for sky estimation are $\sim$20 times the half-light radius 
of target galaxy estimated by SExtractor. 
Top~left: simulated image of a target galaxy (center) and nearby 
objects. 
Top~middle: removing the pixels contaminated from galaxy light 
by $3\sigma$ clip in SExtractor. 
Black contours represent the removed pixels by the $3\sigma$ clip. 
Top~right: histogram of pixels after removal and best-fit 
Gaussian profile where mean and standard deviation are estimated from 
surviving pixels. 
Bottom~left: histogram of pixel values of the true sky background level (black),  
background noise (blue), pure galaxy profiles before sky noise is added to the 
simulation (green), and galaxy profiles after sky noise is added to the simulation (red). 
The true sky surface brightness input into the simulation ($\mu_{\rm sky}~=
~3240~{\rm ADU~pixel}^{-1}$) is subtracted from all curves.
Bottom~middle: spatial filtering of sky pixels 
by masking detected objects out with our masking procedure.
Ellipses indicate our new masks.   
Bottom~right: 
statistical filtering by fitting a Gaussian to histogram of pixels and 
subtracting the extragalactic background light (EBL). 
Solely using the new masking scheme improves upon the background estimate 
from using the SExtractor masks, but continues to overestimate the background. The 
persistent overestimation is resolved, when we subtract the EBL in addition to using 
new masks. The residual in estimated sky value is then 0.011 ADU~pixel$^{-1}$ 
which corresponds to 0.0003\% of the sky.}
\label{skyestimation}
\end{figure*}

In Figure \ref{skyestimation} we compare the performance of background estimation 
by SExtractor and our method for a particular galaxy. It is known that SExtractor 
tends to overestimate sky background \citep{Haussler:2007aa}, and the 
estimate becomes worse when the number of unidentified objects increases. Although 
the 3$\sigma$ clip by SExtractor described in Section~\ref{SExSky}
removes excessively bright pixels from 
consideration when estimating the sky background, the remaining pixels are still
contaminated by the flux from undetected galaxies and insufficiently masked galaxies. 
This can be seen directly in the top right panel in Figure \ref{skyestimation}, which 
shows the distribution of unmasked pixel values when using SExtractor's detection 
masks. While the distribution is well described by a Gaussian, the mean value of the 
distribution is 1.23 ADU~pixel$^{-1}$ above the true sky value. This bias results from 
the excess number of pixels in the bright wing of the distribution. 
In the bottom right panel of Figure \ref{skyestimation}, we show the distribution of 
unmasked pixel values after using our new masking scheme. The bias from extended 
galaxy profiles, which otherwise survives SExtractor's masking procedure, is mitigated 
with our new masks, as the mean value is $\mu$ = 0.910 ADU~pixel$^{-1}$ above the 
true sky value. However, undetected galaxies continue to pollute the pixels used for 
sky estimation with excess flux, even when new masks are used. To deal with flux 
contamination from undetected galaxies, we must also subtract the calculated EBL flux 
per pixel to obtain an accurate estimation of the sky background 
($\mu$ = 0.01 ADU~pixel$^{-1}$). The precision of the estimate can be 
increased by utilizing many such samples of sky over a much larger 
area and requiring smoothness. For this we assume the sky underlying all galaxies 
varies on scales much larger than galaxy scales.\\

\subsection{Global Background Model Coupled With New Local Sky 
Estimator} \label{stack+inchan}
In this sub-section, we describe the procedure to create a global background model for 
entire simulated images. The background model is created by computing the average 
background value in semi-local uniformly spaced subsections of the 
CCD image, and subsequently fitting a smooth two-dimensional 
polynomial to the average background values. Once created, the global background 
model can be evaluated at any point in the image to produce a sky estimation 
\citep{Bosch:2018aa}. Note, this is in contrast to the estimation techniques detailed 
above. Aforementioned techniques are \textit{local} estimators of the sky background: 
using pixels from the surrounding $\sim$ 1 arcminute diameter of a galaxy to construct 
an estimate of the sky brightness in the immediate vicinity of each galaxy. 
The details specific to creating the global background model (subdividing the CCD, 
computing the average background values, and polynomial fitting) are discussed 
below.   

We begin by sub-dividing the image into 2500 evenly spaced, equally sized image 
subsections, where each subsection is 160 $\times$ 160 pixels in size ($\sim$ 40 
square arcseconds), and image subsections do not overlap. The centers of these 
subsections define a 50 $\times$ 50 point spatial grid. In each image subsection, 
an iterative 3$\sigma$ clip mean and variance 
are computed on all pixels which do not correspond to detected objects (i.e.,~the 
pixels not included in the corresponding SExtractor segmentation maps). 
The sigma-clipped mean of each image subsection is assigned to its 
corresponding grid point. The average position of the non-masked pixels in each 
image subsection is used to place the points for the spatial grid in their respective 
image subsections. Subsequently, a 6th-order two-dimensional Chebyshev polynomial is 
fit to the spatial grid. 
Chebyshev polynomials are more robust to over fitting than spline interpolation, as 
they are not strictly required to pass through the grid points obtained from the 3$
\sigma$ clip. Each grid point is inverse-variance-weighted in the fit, so image 
subsections where many pixels are masked have a reduced impact on the fidelity of 
the fit \citep{Bosch:2018aa}. The fitted polynomial model may be used as a 
background model which may be evaluated at any location in the image to predict 
the local sky value.

\begin{figure*}[t]
\centering
\includegraphics[width=0.7\textwidth]{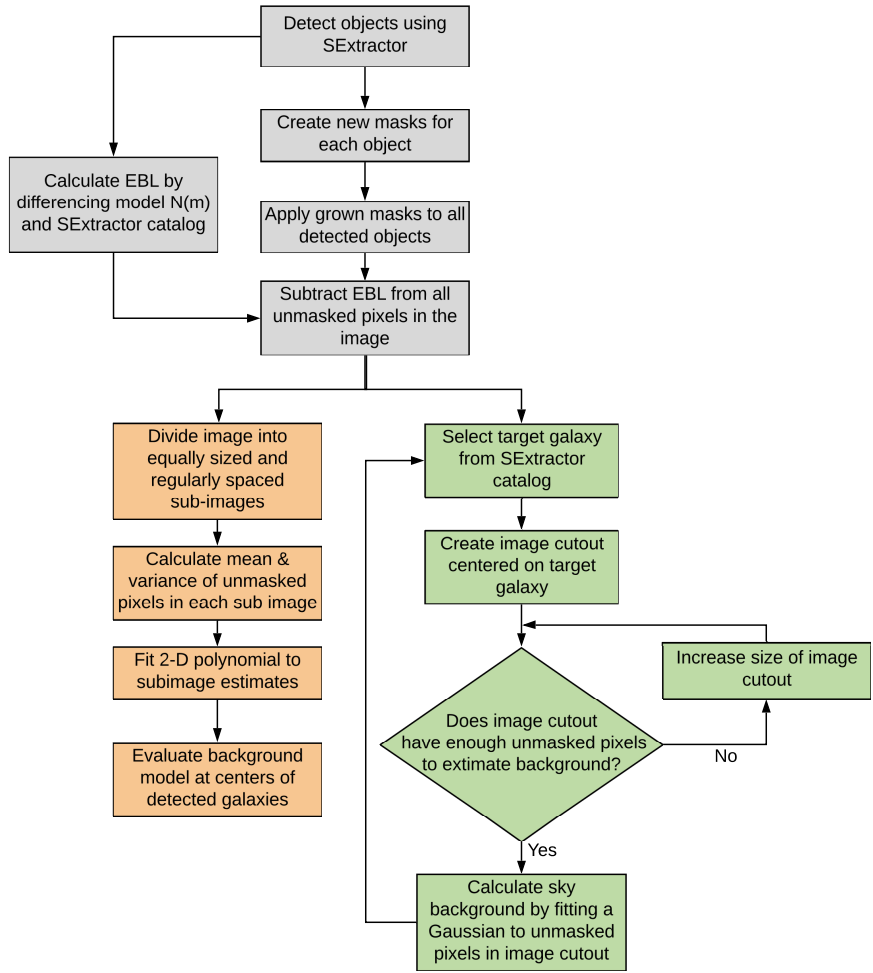}
\caption{
A flowchart describing our two sky estimation processes
as described in Section 4.3 (local estimation) and Section 4.4 (polynomial fitting).
The five gray shaded steps are common to both techniques, while the
color boxes are specific to local estimation (green) and polynomial fitting (orange).
We begin by calculating the residual difference between our model $n(m)$ and
SExtractor's galaxy number count to obtain a statistical estimation of
the EBL. Larger masks are generated and applied to
all detected galaxies. The EBL level is then subtracted from all unmasked pixels. 
The techniques diverge at this juncture. In the local estimation (green) steps,  
image cutouts are generated for a target galaxy, and grown as needed 
until they contain sufficient unmasked pixels. A Gaussian is then fit to remaining
unmasked pixels to estimate the sky for this galaxy. This is repeated for 
all detected galaxies. The polynomial fitting technique (orange) 
creates many equally sized and spaced image subsections over the entire image, 
calculates local means in each subsection, 
and fits 2D polynomial to create a model background. 
}
\label{Flowchart}
\end{figure*}

Several factors must be considered when choosing the image subsection 
size, as this will determine spatial scales the background model is sensitive to. 
If the image subsections are too small, they will include too few surviving unmasked 
pixels from which to calculate a mean. Additionally, if the subsections are 
smaller than the spatial scales of galaxy profiles, 
extended features in galaxy profiles risk being subtracted out. However, all 
other variations in the sky must happen at spatial scales lower than that of 
the sky model if they are to be fitted and removed. Because we have 
restricted our focus to the general case of an extragalactic sky superposed 
on a slowly varying foreground sky level for this work, we have chosen image 
subsection sizes that yield a background model 
sensitive to spatial variations on the order of 40 square arcseconds. 
Investigations in varying the subsection size showed the model is insensitive to 
varying the subsection size between 25 square arcseconds and 1 square 
arcminute. Using subsections smaller than 25 square arcseconds created 
subsections which were completely masked and had no usable pixels from which 
to estimate statistics, and subsections larger than 1 square arcminute 
are sufficiently large to avoid over subtracting extended tails of the 
galaxies simulated here. A summary flow chart of our methods is shown 
in Figure~\ref{Flowchart}.

The appropriate polynomial order for the background model must also be chosen with care, and will depend on a variety of factors. The background in data taken with real cameras on telescopes (in contrast to our idealized simulated data here) will contain contributions due to the optics, like scattered light and diffuse ghosts. Artifacts in detectors, such as tree rings, can give rise to variations across the CCD itself. The spatial scales over which these contributions occur will vary from instrument to instrument and telescope to telescope. This is also true for the astrophysical background itself, as the scales over which it varies can depend on the field of view. The model must vary on spatial frequencies at least as small as those discussed immediately above. If the model varies on scales finer than this,  then the model will be susceptible to over-fitting noise, or fitting and subtracting extended features in galaxy profiles. As a result, the optimal polynomial order in our scheme will depend largely on the data being considered.

Fitting a polynomial background is a somewhat heuristic
procedure when the true underlying sky model is not explicitly known, as is the case for most real observational data. As a consequence, the order of the polynomial used in the background model can be somewhat ad hoc. In \citet{Bosch:2018aa}, the authors use the same polynomial background model discussed above, and find a 6th order polynomial is well suited for modeling the background on the 4K $\times$ 2K CCDs used on the Subaru Hyper Suprime-Cam over the appropriate scales.  Because our simulated data is meant to emulate DLS data, we directly turn to DLS data to investigate the appropriate polynomial order.  We find that while such a model can suffer over subtraction in the immediate vicinity of bright field stars, it otherwise leaves galaxy profiles intact. Our idealized simulations include only galaxies, freeing us from this potential issue. We do not include a spatially varying component of the sky background in our idealized simulation, but for consistency with the methods discussed above, we employ a 6th order polynomial fit in our simulations.

\begin{figure}
\centering
\includegraphics[width=0.47\textwidth]{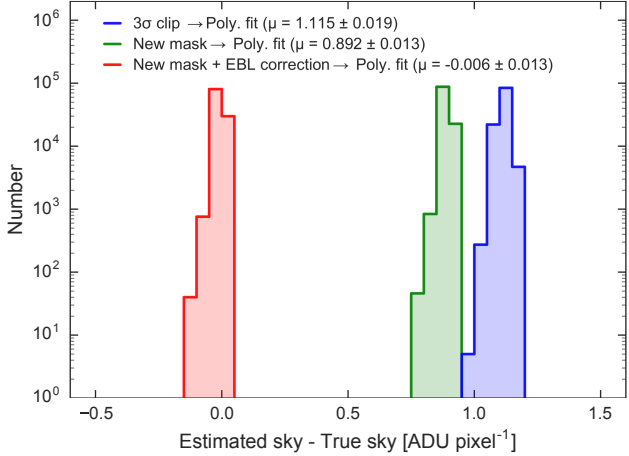}
\caption{Three polynomial background models evaluated at the centroids of 
detected galaxies: with aggressive masking and EBL correction (red), only 
aggressive masking with no EBL correction (green), and SExtractor detection 
map masking with no EBL correction (blue). We use a combination of image 
binning, statistical estimation of sky and its variance in each image bin, 
and polynomial fitting to predict sky background levels at the centers of 
detected galaxies. For each background model, we subdivide the random distribution 
simulation into 160 $\times$ 160 pixel image subsections, and use a statistical 
estimator on all unmasked pixels in each subsection. A spatial 2D polynomial model is 
then fit to the local mean backgrounds to create a global background model, which can 
be used to evaluate the background at the centroids of detected galaxies. 
3$\sigma$ clip with SExtractor detection masks (blue) suffers overestimation 
from extended unmasked galaxy features, and unmasked faint galaxies. Using a 
simple mean and our new masks (green) improves the estimation, but still suffers from 
bias. Only after using both new, larger masks and the EBL correction (red) can accurate, precise sky estimations be made.
}
\label{improved_stack}
\end{figure}

In Figure \ref{improved_stack}, we show the result of this background estimator when 
used on our DLS depth random distribution simulation. In the blue histogram, we show 
the distribution of background estimations at the centroids of detected galaxies. The 
mean of this distribution is 1.115 $\pm$ 0.019 ADU~pixel$^{-1}$ above the true sky 
value. We argue this overestimation is due to flux in the extended, unmasked outskirts 
of galaxy profiles, and undetected galaxies. We attempt to remove this bias using 
techniques described in the previous subsection. The overestimation in sky 
background can be partially ameliorated by recreating the background polynomial 
model, where we use our new masks in lieu of the SExtractor masks. Additionally, a 
simple mean of the unmasked pixels is used to estimate the sky in each image 
sub-section, instead of the 3$\sigma$ clip. In the green curve in Figure 
\ref{improved_stack}, we show the distribution of background estimations at the 
centroids of detected galaxies using this technique. The mean sky value, 
0.892 $\pm$ 0.013  ADU~pixel$^{-1}$, while an improvement 
to the 3$\sigma$ clip method, still suffers from an overestimation. As before, we can 
remove the persistent bias by subtracting the EBL flux in addition to dealing with 
previously unmasked extended galaxy profiles. In the red curve in Figure 
\ref{improved_stack}, we subtract the EBL flux level, estimated from the residual between the true 
and measured galaxy number counts, from each unmasked pixel in the random simulation.
We then repeat the previous procedure to create the green distribution. 
The resulting distribution has a mean of -0.006 $\pm$ 0.013 ADU~pixel$^{-1}$. 
As discussed in Section \ref{newestimationtechnique}, the biases from extended galaxy 
profiles and undetected galaxies must be dealt with in order to create accurate 
estimations of the sky background.

Note, in its implementation in Section \ref{newestimationtechnique}, our sky estimator 
initially takes an image cutout centered on a galaxy, and grows the image 
cutout by 10 pixel increments in width and height as needed to ensure at 
least 4000 unmasked pixels reside in the image cutout. For the global background 
models discussed in this sub-section, however, we do not grow the 
160 $\times$ 160 pixel sub-image when determining the mean and variance 
pixel value.  This is done to ensure that an equally spaced grid is used in 
the polynomial fit, and that the statistics computed for each sub-image are 
representative of the pixels in that sub-image alone.

By combining the new sky estimation technique advocated here with a global 
polynomial background model, we can benefit from ``the best of both worlds.'' 
The polynomial background model captures and smooths over spatial 
fluctuations on the $\sim$ 40 arcsecond scale. This ultimately yields a low 
variance in the distribution of predicted background values at the centers 
of galaxies. The combined model also benefits from the accuracy of our new 
sky estimation technique by accounting for flux contributions from 
undetected sources, and by excluding flux from tails of galaxy profiles 
from detected sources by growing our detection masks.

\begin{figure*}[t]
\centering\includegraphics[width=0.7\textwidth]{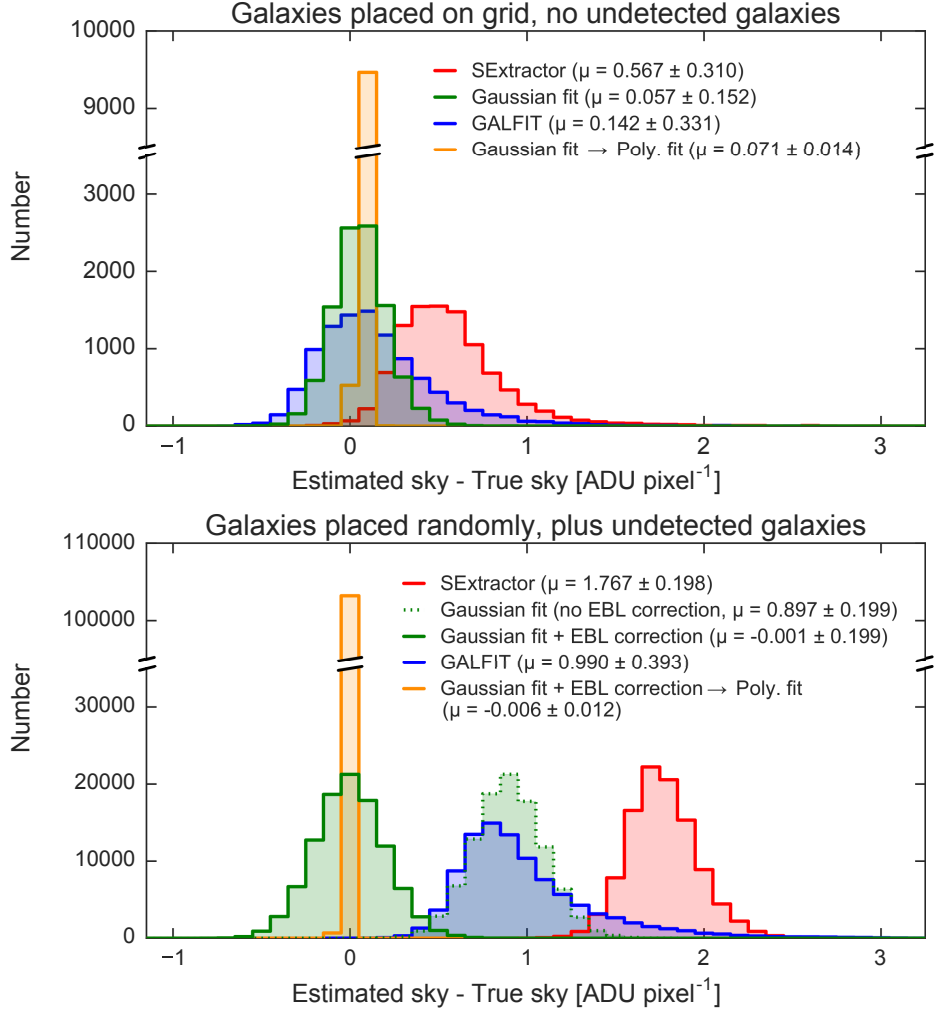} 
\caption{
Top: histogram of estimated sky background in the uniform galaxy distribution 
simulation to DLS depth 
with no neighboring fainter galaxies. The colors of each histogram represent the 
different fitting methods. Means and standard deviations of the difference 
measurements are presented. A difference in the sky background of 
1 ADU~pixel$^{-1}$ corresponds to 0.03\% error in terms of sky surface brightness. 
For this simulation, we fit a Gaussian to data without modeling the EBL 
because we have detected all galaxies. If detected galaxies are well masked 
out, fitting a Gaussian performs better than GALFIT and SExtractor. 
Bottom: histogram of estimated sky background in the random galaxy 
distribution simulation to DLS depth, 
which includes much fainter galaxies (EBL). 
Fitting a Gaussian without accounting for the EBL to data 
overestimates sky background (green dotted line). However, combining a 
Gaussian with the EBL statistical estimate recovers the true values 
of sky background (green solid line).  Furthermore, the use of a polynomial 
fit with EBL corrected local estimates 
reduces dispersion of sky estimation to 4 ppm of the sky level. A  bias in sky 
estimation propagates to a systematic offset in measured photometry, and our 
method reduces this bias by more accurately recovering the true sky level. }
\label{sky:set}
\end{figure*}

\subsection{Comparing different sky estimators}
We compare sky background estimates for detected galaxy images in our DLS-depth simulated images. In 
the uniform distribution simulation, all galaxies are detected and cataloged; 
consequently, we use our estimators on all galaxies in this simulation. 
In the random distribution simulation, however, we only consider galaxies which 
SExtractor detected and for which GALFIT is able to converge on a sky estimation. 
The numbers of galaxies that meet the criteria above are 10,000 and 98,149 
for the uniform and random simulations, respectively. In Figure \ref{sky:set}, we 
compare sky estimations by SExtractor, GALFIT, Gaussian fit + new masks (with and 
without EBL correction), and polynomial background model
with EBL correction and new masks. Note that the true sky background is 3240 
ADU~pixel$^{-1}$ and this pedestal has been subtracted in the uniform and random 
distribution simulations, although the shot noise from this sky level is included.

The ensemble of SExtractor local background estimations has a mean of 0.567 
$\pm$ 0.310 ADU~pixel$^{-1}$ and 1.767 $\pm$ 0.198 ADU~pixel$^{-1}$ for the 
uniform and random distribution simulations, respectively. As we discussed in 
Section \ref{GALSIM}, each image cutout in the uniform distribution simulation is 
completely isolated from the flux of neighboring galaxies, allowing us to test 
background estimation independent of blending and crowding effects. Even so, 
SExtractor overestimates the sky background for simulated galaxies in the uniform 
distribution. This is due to the flux residing in the extended galaxy profiles SExtractor 
fails to mask. If the measurement is done in a crowded region with 
a number of undetected galaxies (the random distribution simulation), the 
overestimation is compounded by the flux of these undetected galaxies. 

GALFIT background estimates are better than SExtractor background estimates; 
GALFIT estimates average local backgrounds of 0.142 $\pm$ 0.331 ADU~pixel$^{-1}$ 
and 0.990 $\pm$ 0.393 ADU~pixel$^{-1}$ for the uniform and random 
distribution simulations, respectively. The distributions of GALFIT 
background estimates have noticeable extended tails, and comparatively 
larger scatters than other methods. This is likely due to GALFIT's sensitivity to noise. 
Nevertheless, the peak values in the pixel histograms as shown in Figure \ref{sky:set} 
are close to the true value in the uniform distribution simulation and the EBL 
in the random distribution simulation.

Estimation of sky background in our hybrid method has the highest precision 
of all in terms of mean and standard deviation of histogram, and is immune to the 
environments we have tested. Before using the polynomial fit, average local backgrounds are  
0.057 $\pm$ 0.152 ADU~pixel$^{-1}$ and -0.001 $\pm$ 0.199 ADU~pixel$^{-1}$ 
for the uniform and random distribution simulations, respectively. As we 
discussed in Section \ref{newestimationtechnique}, we can reduce the noise in sky 
estimation by applying the polynomial model to the sky estimates obtained 
from fitting a Gaussian profile to the pixel distribution in each image 
cutout. The resulting sky backgrounds (after EBL correction for the random 
distribution simulation) are 0.071 $\pm$ 0.014 ADU~pixel$^{-1}$ and -0.006 
$\pm$ 0.012 ADU~pixel$^{-1}$ for the uniform and random distribution 
simulations, respectively. By doing so, we reduce the uncertainty of sky 
estimation by a factor of 10 or more; most sky estimates lie within 
$\pm$ 0.0004\% (0.015 ADU~pixel$^{-1}$). 
Below we investigate the precision of recovery of surface 
brightness profiles of simulated galaxies.

\section[t]{S\'e{r}sic Index Recovery with Various Sky Estimators}
\label{sersicest}
In this section we focus on the effects of sky estimation accuracy on the apparent 
galaxy morphology. Since Edwin Hubble's first study of galaxy classification using their 
appearance \citep{Hubble:1926aa}, connections between galaxy morphology, shape, 
and color have provided insight into galaxy formation and 
evolution. Further methods for galaxy classification have been developed using 
the one-dimensional radially averaged profile of galaxy surface brightness  
\citep{de-Vaucouleurs:1948aa, Sersic:1963aa}. Among several functional forms for 
the profiles, the S\'{e}rsic profile is one of the most popular. The S\'{e}rsic 
profile is a fitting function that describes the surface brightness profile 
(the intensity of light as a function of distance from the center) given by: 
\begin{equation}
\Sigma(R) = \Sigma_e \exp\bigg[-b_n\bigg\{ \bigg( \frac{R}
{R_e}\bigg)^{1/n} - 1 \bigg\}\bigg]
\end{equation}
where $\Sigma_e$ is the effective intensity (the surface brightness at the 
effective radius), $R_e$ is the effective radius which encloses half of the 
total light (half-light radius, hereafter), $b_n$ is the concentration which is 
defined so that half of the total light is inside the half-light radius for a 
given S\'{e}rsic index \citep{Graham:2005aa}, and $n$ is the S\'{e}rsic index 
which describes the shape of profile and is correlated with galaxy surface 
brightness morphology.

As an illustrative example of the impact different background estimation techniques 
can have on source measurement, we measure the surface brightness profiles of 
model galaxies after subtracting the sky background using the three techniques 
discussed in Section \ref{skyestimators}: SExtractor, GALFIT, and our new method.  
We measure the S\'ersic index in all simulations using GALFIT.  As galaxy profile fits 
are sensitive to flux from nearby objects that are blended with the galaxy of interest, 
we restrict this study to the uniform distribution simulations where blending is not an 
issue to isolate the sky subtraction effect from the blending effect.  We do so for both 
DLS and LSST DDF depths. To no surprise, we find that the shape of the faint outer
surface brightness tail is affected by sky level mis-estimation. 

The GALFIT output parameters are very sensitive to the parameters in the GALFIT 
start file.  For this test, we use parameters derived from SExtractor. The initial 
parameters are determined as follows:  the size of image cutout is the same as used 
for sky estimation (see Section \ref{skyestimators}). The total magnitude is given by 
\verb+MAG_AUTO+; the half-light radius is given using  \verb+FLUX_RADIUS+ with 
\verb+PHOT_FLUXFRAC+ = 0.5. The axis ratio $b/a$ and the 
position angle are derived by taking 1 - \verb+ELLIPTICITY+ and 
\verb+THETA_IMAGE+, respectively. A S\'ersic index of $n$ = 2.5 is assumed as 
an initial guess in the fitting. For the PSF convolution, we generate the PSF image 
using GALSIM with the true parameters after matching the SExtractor's position and 
true position. 
  
\begin{figure*}[t]
\centering
\includegraphics[width=0.7\textwidth]{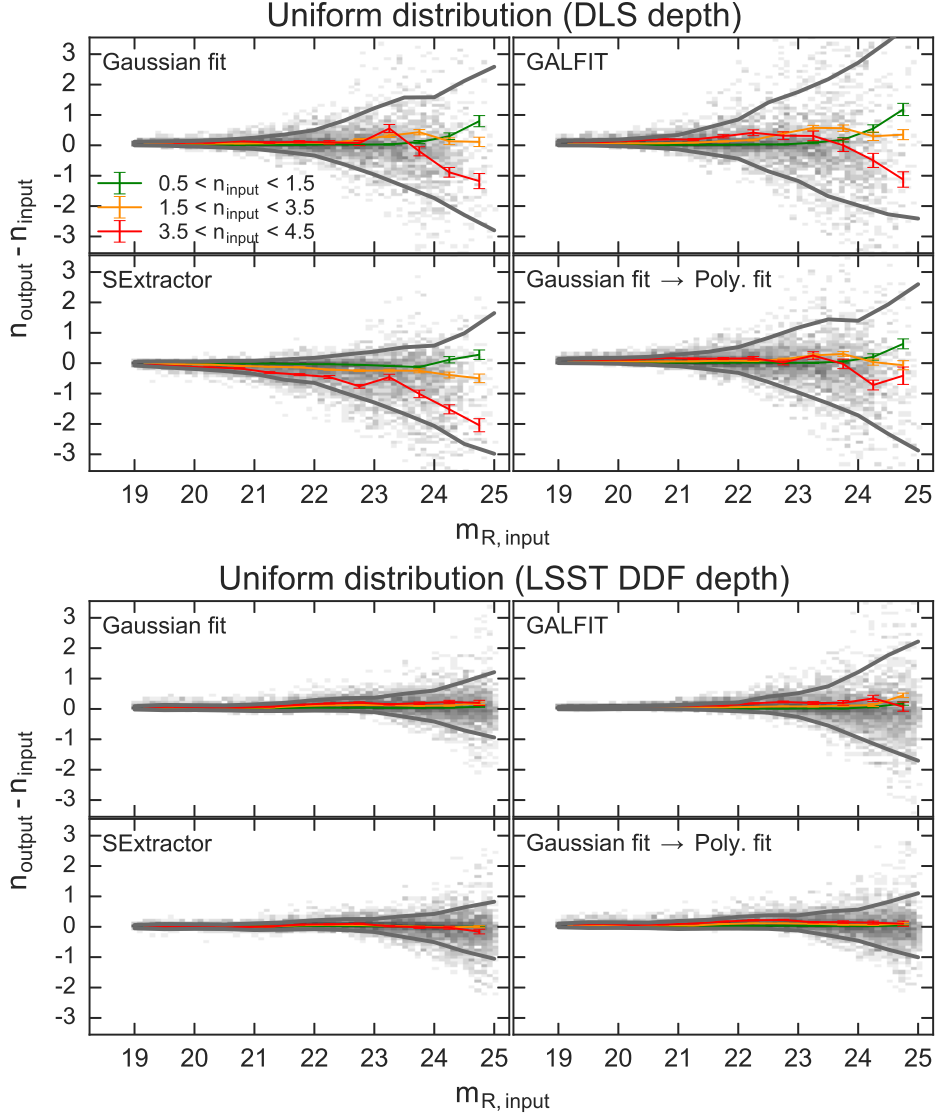}
\caption{
Difference between the true and estimated S\'ersic index as a function 
of magnitude for the uniform galaxy 
distribution simulation to DLS depth (Top) 
and LSST DDF depth
(Bottom). Gray 
scale indicates the number density and  gray thick lines represent 
$\pm1\sigma$ values with respect to mean for different magnitude bins.
Colored solid lines represent the mean difference between input 
S\'ersic index and our estimated S\'ersic index for different input 
S\'ersic index bins: 
$0.5<n_{\rm input}<1.5$ (green), 
$1.5<n_{\rm input}<3.5$ (orange), and
$3.5<n_{\rm input}<4.5$ (red). 
Note that each image cutout contains only one galaxy. 
GALFIT runs combined with our sky estimator recover  
the true S\'{e}rsic indices in both simulations down to the noise floor. 
In addition, the accuracy of S\'ersic index estimation at LSST DDF depth is  
much greater than at DLS depth. This is a purely S/N effect, as high S\'ersic 
index galaxy fits are sensitive to the noise level at large radii.
}
\label{sersic:mag_n}
\end{figure*}

We investigate the effect of sky background on determining galaxy surface brightness 
profile types. In Figure \ref{sersic:mag_n} we compare the measured S\'ersic index 
versus the true value of S\'ersic index using different sky estimators for both the 
DLS-depth and DDF-depth uniform simulations. We compare all galaxies with which  
GALFIT fits do not fail nor produce problematic parameters. Galaxy surface 
brightness fits do not converge for fainter galaxies because their sizes become small, 
and the image cutout becomes noise-dominated. The numbers of galaxies with good 
fits are 8,428 and 9,784 for DLS and LSST DDF depths, respectively.

Overall, our estimation of sky background performs more robustly than the other 
methods. We note that the scatter of our technique increases for large S\'ersic index. 
Noise fluctuation in the faint, extended tails of large-S\'ersic-index galaxies negatively 
impacts the fidelity of profile fitting. By contrast, the surface brightness profile of 
low-S\'ersic-index galaxies drops rapidly, so galaxy surface brightness profile 
estimations are less affected by noise in these cases.

We also investigate the dependence of the S\'ersic-index estimation 
on the S\'ersic-index and magnitude of galaxies. For a given sky estimate, 
sky background and S\'ersic index are anti-correlated: for sky estimators 
that tend to overestimate the background, there is a corresponding underestimate 
of S\'ersic-index. This is more clearly seen as the brightness decreases or 
S\'ersic-index increases. This trend is also found in previous survey data like 
the SDSS where sky background is overestimated 
\citep{Blanton:2005aa,Blanton:2011aa}. There is an obvious explanation as to why 
there is an overestimate of S\'ersic-index for high-S\'ersic-index galaxies: 
in these cases, the true galaxy profile tends to be truncated at large radii due to the 
over-subtraction of sky background. 

In our DLS-depth uniform simulations, our new sky background estimator results in an 
unbiased and less-scattered estimation of S\'ersic-index. For bright galaxies 
(i.e., $m_{\rm R}\le$ 24 for our study), there is no discernible trend in over- or 
under-estimation of galaxy surface brightness morphology, i.e. S\'ersic-index. 
For fainter galaxies, however, uncertainties in the fits increase, and a strong 
dependence on galaxy surface brightness morphology is seen. 
Nonetheless, our new estimator still outperforms the other sky estimators in this 
regime.

The performance of GALFIT runs using different 
sky estimators are less distinguishable at LSST DDF depth. 
Unlike DLS depth, the mean values of estimated S\'ersic indices 
match the true values for all GALFIT runs. No bias is found as 
the magnitude becomes fainter or as the true S\'ersic index increases. 
This is because the sky noise in LSST DDF depth is much smaller than 
DLS depth. Also, there is no discernible difference between our work 
(Gaussian fit and Gaussian fit $\rightarrow$ Poly. fit) and SExtractor.
However, the GALFIT runs are still noisier than the other methods. 
This is due to the fact that the sky value estimated by GALFIT 
($\mu_{\rm sky}=-0.049\pm0.225$) has larger error than the other methods 
($\mu_{\rm sky}=0.031\pm0.022$, 0.011$\pm$0.011, and 0.014$\pm$0.002 
for SExtractor, Gaussian fit, and Gaussian fit $\rightarrow$ Poly. fit, 
respectively). 
We conclude that for the isolated galaxy case signal-to-noise ratio is a major 
factor in the surface brightness profile estimation. 
For example, see \citet{Taghizadeh-Popp:2015aa}, who show an underestimate 
of galaxy size near the detection limit at multiple depths. This truncation in size 
is closely related to the bias in profile estimate.
Good signal-to-noise 
is a necessary but not sufficient condition for accurate profile fitting 
when faint undetected galaxies are included. 
As we found before in the DLS depth case, accounting for 
the undetected galaxies corrects for this sky level bias, leading to more 
accurate outer profile fits.  However, the presence of unresolved background 
galaxies disturbs the S\'ersic-index fit for individual galaxies even at fairly 
high signal-to-noise, leading to increased bias and scatter for all estimators 
tested compared to the uniform simulation case, where no confounding 
galaxies are present 
nearby. We will explore these effects in more detail in a future paper.

\section{Discussion and Future Work}
\label{discussion}

In this paper we present sky background estimation using various publicly 
available packages, and compare with results using our new method that grows the 
spatial masks around detected objects, and statistically accounts for flux from 
undetected faint galaxies. Our algorithm is able to recover the sky level to 
4 ppm in our simulated data, an improvement over existing sky background 
estimation techniques. Our analysis is confined to simulations of the extragalactic 
sky components with added spatially uniform sky foreground;  optics ghosts and 
scattered light around bright stars are beyond the scope of this paper.

We demonstrate that insufficiently masking the extended features of 
galaxies can bias sky estimation high. This occurs because flux from these 
galaxy regions contaminate the pixels used to estimate the sky background. 
While widely used estimators suffer from this bias, our technique is able 
to overcome it; by conservatively masking galaxies, our estimator considers 
pixels which are truly more representative of the true sky background.
We also show that successful estimation of the underlying sky background must 
consider flux contributions from undetected faint background galaxies. 
To correct this bias, we use knowledge 
of galaxy counts as a number of magnitude to accurately estimate the flux contribution 
from these galaxies.

To demonstrate the power of our new technique, we obtain galaxy surface 
brightness profile fits, via the S\'ersic index, using different sky estimators. Previous 
methods overestimate sky background, resulting in incorrect S\'ersic estimates, and all
show large scatter. In contrast, our two-filter estimator has the highest 
precision and is least affected by simulation details such as the brightness, the surface 
brightness profile, and the galaxy number density.

While it is beyond the scope of this paper, there are additional steps that may be taken 
to improve our technique, so that 
real imaging data may fully benefit from it. In our demonstration of this hybrid 
sky estimation algorithm we have used a simplified galaxy number count 
distribution, $n(m)$. For a more realistic approach in applying this sky estimator 
to real data one should use a $n(m)$ slope that becomes shallower 
beyond  $m_{\rm R} \simeq$  26 to more realistically represent the observed faint 
end of magnitude distribution. Additionally, we do not simulate internal reflections and 
scattered light in the camera, or other sources of sky variation, all of which will have to 
be adequately modeled for each exposure in real data. Ultimately, the level 
of accuracy offered by our technique is a necessary but not sufficient 
condition for exploration of ultra-low surface brightness.  Our simulation placed galaxies 
randomly in the image plane; however, real galaxies are embedded in large scale 
structures, and cluster with each other.  This may lead to a slight overdensity of 
undetected EBL galaxies in the vicinity of detected objects that is not reflected in 
our random simulation.  Such an excess could still pollute the outer isophotes of 
the galaxy profile. The large number of undetected galaxies over a wide range of 
redshifts in projection strongly mitigates this effect.

We emphasize that developing detection algorithms operating at 
ultra low surface brightness which take full advantage 
of such high precision sky estimates is beyond the scope of this paper, though such 
sky precision would be a prerequisite. An added challenge is fitting sky across CCDs in a mosaic.
For this, we note that the LSST Project has recently implemented a superior sky estimator which 
fits the background over an entire exposure.\footnote{https://community.lsst.org/t/sky-subtraction/2415}
This allows using a larger scale for the background model, including the removal of static structures
(such as the average response of the camera to the sky) in the background.

The accuracy of fitting galaxy surface 
brightness profiles will be improved with multi-wavelength photometry since 
structural parameters of galaxies vary in different photometric bands, though 
they are correlated \citep{Hausler:2013aa}. In such a joint fit our sky 
estimator can contribute to enhanced profile accuracy, particularly in low 
signal-to-noise bands.
Optimal detection depends on the angular scale of the object: therefore
the underlying sky background must be defined on that scale and larger scales. 
Thus, scale dependent sky models must be developed which include all components 
of the apparent sky on relevant scales, from telescope optics ghosts to large scale dust.

There are many science 
drivers which rely on detection of low surface brightness features. Increased 
precision in the measurement of sky background can be 
applied to a better understanding of the evolution of galaxy surface 
brightness morphology \citep{Conselice:2003aa,Conselice:2014aa}.  One can also 
probe the evolution of mass structure, surface brightness profile, and star 
formation of galaxies as a function of redshift with less bias at low 
surface brightness. Studies of the dark halo stellar halo connection would 
be less biased: there is a correlation between dark matter structure and 
light distribution of galaxies at late cosmic time \citep{Wetzel:2015aa,Huang:2017aa,Somerville:2017aa} 
because galaxies have different star formation histories depending on 
stellar mass \citep{Qu:2017aa} and galaxy surface brightness morphology 
\citep{Wuyts:2011aa}. Finally, it is possible, and even likely, that unexpected 
discoveries lie at low surface brightness levels.
Depending on the angular scale of the object being studied, such 
applications of sky estimation will require a full multi-component model of 
the apparent sky. The most demanding application is the unbiased detection 
and photometry of ultra LSB galaxies of large half-light radius. 
Using a noise-based non-parametric technique may be a better approach 
to detect such ultra faint sources. The faint-source detection capability resulting 
from this method has shown improved results for faint sources relative to the 
signal-based source detection algorithm employed by SExtractor \citep{Akhlaghi:2015aa}.  
For most 
of these cases, a robust sky estimation with accuracy of a few parts in 
$10^5$ or better, and at high precision is required.

\section*{Acknowledgements}
We thank Perry Gee, Lee Kelvin, Robert Lupton, Chien Peng, Brant Robertson, and Paul Price for 
helpful discussions. The sky sample fitting algorithm which we use is part of 
the sky estimator in the 2017 LSST Stack, for which we acknowledge the efforts of 
Steve Bickerton and Russell Owen. We thank Andrew Bradshaw and Craig Lage for 
comments on the manuscript. We thank the referee for suggestions that improved the 
manuscript.  Support from DOE grant DE-SC0009999 
and NSF/AURA grant N56981C is gratefully acknowledged.

\end{document}